\documentclass[a4paper,11pt]{article}
\usepackage{pos}
\usepackage{subcaption}
\usepackage{graphicx}
\usepackage{float}

\def\thefootnote{\fnsymbol{footnote}} 
 
\usepackage{comment}
\renewcommand{\logo}{\relax}

\newcommand{\MSbar}{{\overline{\mbox{MS}}}} 
\newcommand{\MSbars}{\overline{\mbox{\footnotesize{MS}}}} 
\newcommand{\mMOM}{\mbox{mMOM}}
\newcommand{\mMOMs}{{\mbox{\footnotesize{mMOM}}}}

\newcommand{\RI}{\mbox{RI${}^\prime$}}

\newcommand{\Nf}{N_{\!f}}

\newcommand{\Deltacsb}{\Delta_{\mbox{\small csb}}}

\renewcommand{\hookAfterAbstract}{%
\par\bigskip
\textsc{ArXiv ePrint}:
\href{https://arxiv.org/abs/2309.16554}{2309.16554}
}

\begin{document}
\title{Crewther’s relation in different schemes}
\author*{Robert H. Mason}
\author{John A. Gracey}
\affiliation{Theoretical Physics Division,\\ Department of Mathematical Sciences, University of Liverpool, Liverpool, United Kingdom}
\emailAdd{robert.mason@liverpool.ac.uk}
\emailAdd{gracey@liverpool.ac.uk}
\abstract{We examine Crewther’s relation at high loop order in perturbative QCD and demonstrate how the relation is accommodated in gauge-parameter dependent schemes where the running of the gauge parameter has to be explicitly considered. Motivated by ensuring that the conformal properties of the relation are preserved at all the critical points of QCD, including the Banks-Zaks and its infra-red stable twin, we demonstrate the necessity of an additional term in the relation for describing gauge running in the minimal momentum subtraction scheme (mMOM) and argue for its inclusion for all gauge-parameter dependent schemes. }
\FullConference{16th International Symposium on Radiative Corrections: Applications of Quantum Field Theory to Phenomenology (RADCOR2023)\\
28th May - 2nd June, 2023\\
Crieff, Scotland, UK
}
\maketitle

\section{Introduction}
Investigations of Crewther's relation have typically been undertaken in gauge-parameter independent schemes where all $\beta$-function coefficients are independent of the gauge parameter, as in $\MSbar$ \cite{1}. The relation, as stated for these schemes, connects two measurable quantities: the Adler D function ($D$) and the Bjorken sum rule ($C$), to the $\beta$-function ($\beta$) and a perturbative series which we will refer to as the Crewther series ($K$). This is described by the equations 
         \begin{eqnarray}
            C(a)D(a)&=&d_R(1+\Deltacsb(a)) \qquad \mbox{where} \qquad \Deltacsb(a)=K(a)\frac{\beta(a)}{a} \label{eq:CrewtherEq},
\end{eqnarray} 
where $a=\frac{g^2}{16\pi^2}$. In \cite{2} it was recognised there were no $\mathcal{O}(a)$ corrections to the product and \cite{3} codified the higher order corrections into the above form. This relation has since been verified in $\MSbar$ to all available loop orders \cite{4} as well as several other gauge-parameter independent schemes e.g. the V-scheme in \cite{5}, and arguments have been made as to its validity to all orders \cite{6,7}. A consequence of this decomposition is that at a fixed point the product reduces to the constant $d_R$. For gauge-parameter independent schemes measurable quantities have a single coupling and therefore fixed points, which we label as $a^\infty$, occur at the roots of the $\beta$-function such that $\beta(a^\infty)=0$, thus by Eq. (\ref{eq:CrewtherEq}) we have $C(a^\infty)D(a^\infty)=d_R$. In practice this can only be said to be true to the order in truncation of the small coupling constant meaning when it is evaluated numerically this will not be an exact result. If the series $K(a)$ were entirely unstructured the statement in Eq. (\ref{eq:CrewtherEq}) could be ensured for any product of two series if one took entirely arbitrary expansion coefficients of the Crewther series. However, additional structure on the coefficients is suggestive of deeper meaning. In particular, they only contain positive powers of the number of fermions $N_f$ as well as of the colour factors, along with other properties described in \cite{3}. 

When applied to gauge-parameter dependent schemes, such as mMOM, this structure failed except in particular choices of the gauge parameter \cite{9,10}.  In this report we review the extension of Crewther's relation to gauge-parameter dependent schemes, suggested in \cite{11}, which adds a second term to $\Deltacsb$ to include the running of the gauge parameter. In Section \ref{sec:SchemeChange} we present and briefly derive this form from renormalization group arguments. Following this in Section \ref{sec:CrewmMOM} we consider the resulting pair of Crewther series for the mMOM scheme (defined in \cite{12}). Section \ref{sec:FixedPoints} is devoted to arguing for the necessity of the additional term through numerical experimentation at the fixed points of gauge-parameter dependent mMOM scheme, provided in \cite{13}. Finally, Section \ref{sec:outlook} provides a discussion of the perspective gained through this investigation of the Crewther relation.

\section{Scheme Change}\label{sec:SchemeChange}
$\Deltacsb$ is related to the product of two measurable quantities by a constant scale and addition and thus is itself a measurable. Its value should therefore be invariant under a scheme change. The $\beta$-function however is not and transforms under a change in scheme by the equation 
   \begin{eqnarray}
\beta_{\MSbars}(a)&=&\beta_{s}(a_{s},\alpha_{s})\frac{\partial a(a_{s},\alpha_{s})}{\partial a_{s}}+\alpha_{s}\gamma_\alpha^{s}(a_{s},\alpha_{s})\frac{\partial a(a_{s},\alpha_{s})}{\partial \alpha_{s}}.\label{eq:betachange}
\end{eqnarray}
where $a$ denotes the coupling in the $\MSbar$ scheme which is gauge-parameter independent and $a_s$ and $\alpha_s$ are the coupling constant and gauge parameter in the target gauge-parameter dependent scheme $s$. The running of the gauge parameter is described with $\alpha_s\gamma^s_\alpha(a_s,\alpha_s)=\frac{d\alpha_s}{dl}$ where $l=\ln\left(\tfrac{\mu^2}{\Lambda^2}\right)$. Consider now rewriting the conformal symmetry breaking term in terms of the couplings of the new scheme from the original form as given in Eq. (\ref{eq:CrewtherEq}) we find
\begin{eqnarray}
\Deltacsb(a)=\Deltacsb(a_s,\alpha_s)=K_a(a)\Bigg|_{\MSbars \to s}\Bigg[\beta_s(a_s,\alpha_s)\frac{\partial a(a_s,\alpha_s)}{\partial a_s}+\alpha_s\gamma_\alpha^s(a_s,\alpha_s)\frac{\partial a(a_s,\alpha_s)}{\partial \alpha_s}\Bigg]
\end{eqnarray}
where we have defined $K_a=\frac{K}{a}$ which is a perturbative series since $K=\mathcal{O}(a)$ and $\MSbar \to s$ implies use of the coupling constant conversion function to change from the coupling of the original scheme to the new couplings of the new scheme. The above equation suggests a new decomposition
        \begin{eqnarray}
\Deltacsb(a_{s},\alpha_{s})&=&K_a^{s}(a_{s},\alpha_{s})\beta_{s}(a_{s},\alpha_{s})+K_\alpha^{s}(a_{s},\alpha_{s})\alpha_{s}\gamma_\alpha^{s}(a_{s},\alpha_{s}),\label{eq:shinyCrewtherEq}
\end{eqnarray}  
whose series can be found through the relations
\begin{eqnarray}
        K_a^{s}(a_{s},\alpha_{s})
&=&\frac{\partial a}{\partial a_{s}}\Big[K_a(a)\Big]\Big|_{\MSbars \to s} \quad \mbox{and} \quad     K_\alpha^{s}(a_{s},\alpha_{s})=
\frac{\partial a}{\partial \alpha_{s}}\Big[K_a( a)\Big]\Big|_{\MSbars \to s} \label{eq:CrewtherSchemeChange}.
    \end{eqnarray}
While these arguments are suggestive that the Crewther relation should be extended to include a gauge parameter running term it does not prove the necessity of the additional term since it could be possible that a decomposition of the form given in Eq. (\ref{eq:CrewtherEq}) is possible for other schemes when purely considering the above derivation. In the proceeding sections we will argue for the necessity of the additional term.

\section{Crewther Series in the $\mMOM$ scheme}\label{sec:CrewmMOM}   
In \cite{10,11} it was found that when the original Crewther relation is applied naively to the mMOM scheme it fails at $\mathcal{O}(a^3)$, except in the Landau ($\alpha=0$), anti-Yennie ($\alpha=-3$) and anti-Feynman ($\alpha=-1$) gauges, although these latter two fail again at $\mathcal{O}(a^4)$. This provides a strong argument against the original Crewther relation for mMOM and therefore for gauge-parameter dependent schemes in general, since we expect physically meaningful relations to be true in all gauges.

With this in mind we begin our discussion of the Crewther series in the mMOM scheme by calculating the $K_a$ and $K_\alpha$ term in this scheme from the conversion functions found in Eq. (\ref{eq:CrewtherSchemeChange}), which using results from \cite{9,10,15,16}, gives
  \begin{eqnarray}
K^{\mMOMs}_a(a,\alpha)&=&16{\zeta_3}-14+\Big[  ( -15552{\zeta_3}+13608 ) {\alpha}^{2}+ ( -31104{\zeta_3}+27216 ) \alpha \nonumber\\
&&+ ( -20736N_f+628416 ) {\zeta_3}+26784N_f-276480\zeta_5-483432 \Big] \frac {a}{648}\nonumber\\
&&\Big[  [ -52488\,{\zeta_3}+45927] {{\alpha}}^{3}+ [ 104976\,{{\zeta_3}}^{2}+ \left( 46656\,{N_f}-1768230 \right) {\zeta_3}\nonumber\\
&&-60264\,{N_f}+622080\,{\zeta_5}+1317357 ] {{\alpha}}^{2}+ [ -769824\,{{\zeta_3}}^{2}\nonumber\\
&&+ \left( 93312\,{N_f}-1740204 \right) {\zeta_3} -120528\,{N_f}+1244160\,{\zeta_5}+1813131 ] {\alpha}\nonumber\\
&&+ ( 207360\,{N_f}-8215344 ){{\zeta_3}}^{2} + (48384\,{{N_f}}^{2}-2782848\,{N_f}+37896114 ){\zeta_3}\nonumber\\
&&+ ( 69120\,{\zeta_5}-112896 ) {{N_f}}^{2}+ ( -2073600\,{\zeta_5}+4804184 ){N_f}+5078400\,{\zeta_5}\nonumber\\
&&+4838400\,{\zeta_7}-43011419 \Big] \frac {a^2}{648} +\mathcal{O}(a^3)
\end{eqnarray}
and
\begin{eqnarray}
K^{\mMOMs}_\alpha(a,\alpha) &=& -24 (\alpha+1) \Big(\zeta_3 -\tfrac{7}{8}\Big)a^2 +  \Big[ \Big( -5832 {\zeta_3}+5103 \Big) \frac{\alpha^2}{72} \nonumber\\
&&+\Big( 7776 {{\zeta_3}}^{2}+ ( 3456 {N_f}-130980 ) {\zeta_3}-4464 {N_f}+46080 {\zeta_5}+97582 \Big)\frac{\alpha}{72}\nonumber\\
&&-396 {{\zeta_3}}^{2}+ \Big( 3456 {N_f}-64452 \Big) \frac{\zeta_3}{72}-62 {N_f}+640 {\zeta_5}+\frac{67153}{72}\Big]a^3\nonumber\\&&+\mathcal{O}(a^4),
\end{eqnarray}
where we have presented the SU(3) expression to reduce the size of the equations, and $a$ and $\alpha$ refer to the coupling in the mMOM scheme. Considering $K^{\mMOMs}_\alpha(a,0)$ we find the same Crewther series provided in \cite{9,10} for the case of the Landau gauge which is as expected because at $\alpha=0$ under linear covariant gauge fixing we can ignore the $\alpha\gamma_\alpha$ term in our Crewther decomposition and therefore we will be left with the $K_a$ term alone. The $\alpha=-1$ gauge can also be understood as the leading term in $K_\alpha$, at $\mathcal{O}(a^3)$ in $\Deltacsb$, has a factor of $\alpha+1$ which disappears in this selected gauge. However, the next-to-leading order $K_\alpha$ term does not have a similar factorisation. So the $K_\alpha$ term cannot be ignored to $\mathcal{O}(a^4)$ in the same gauge. 

If one were to calculate the series $K_a$ and $K_\alpha$ for the mMOM scheme directly from the product and Eq. (\ref{eq:shinyCrewtherEq}) it is not certain you would arrive at the above series. Inspection of the equation tells us there is an ambiguity in the choice of the series where $\bar{K}_a$ and $\bar{K}_\alpha$ could be substituted for $K_a$ and $K_\alpha$ respectively, in Crewther's relation provided they obey the relations
    \begin{eqnarray}
        \bar{K}_a^{s}(F;a_{s},\alpha_{s})&=&K_a^{s}(a_{s},\alpha_{s})-F(a_{s},\alpha_{s})\alpha_{s}\gamma_\alpha^{s}(a_{s},\alpha_{s}),\nonumber\\
        \bar{K}_\alpha^{s}(F;a_{s},\alpha_{s})&=&K_\alpha^{s}(a_{s},\alpha_{s})+F(a_{s},\alpha_{s})\beta_{s}(a_{s},\alpha_{s}).\label{eq:CrewtherAmbiguity}
    \end{eqnarray}
    where $F(a_s,\alpha_s)$ is a generic perturbative series. We note that we have reverted to noting the general scheme as these relations are not specific to the mMOM. If we were able to preserve the original Crewther relation in this schemes then we would require the existence of a series $F_0$ such that $\bar{K}^s_\alpha(F_0;a_s,\alpha_s)=0$. In the mMOM scheme this would require a series such that
    \begin{eqnarray}
    F_0(a,\alpha)=-\frac{K_\alpha^{\mMOMs}(a,\alpha)}{\beta^{\mMOMs}(a,\alpha)}\approx-\frac{24(\alpha+1)(\zeta_3-\frac{7}{8})}{11-\frac{2}{3}N_f}+\mathcal{O}(a).
    \end{eqnarray}
Again we see that this equation could be ensured to each order in general if we were not to enforce additional constraints on the series coefficients such that they are valid perturbative coefficients as was found for the original coefficients in \cite{3}. This suggests we cannot find a perturbative series $K_a$ such that there is no term to describe the gauge running for a general gauge parameter. In the next section we will discuss the importance of the new decomposition through numerical evaluation of the product at fixed points of the running.

\section{Fixed Points}\label{sec:FixedPoints}
Within Crewther's relation we parametrise the conformal symmetry breaking by $\Deltacsb$ which should disappear when the system becomes invariant of the scale of the problem. These points exist when the running couplings of the theory become stationary which for gauge-parameter independent schemes will be at the roots of the $\beta$-function, and for gauge-parameter dependent schemes  we add to this condition the requirement that $\alpha\gamma_\alpha=0$ \cite{14}. By inspection of the original form of this quantity given in Eq. (\ref{eq:CrewtherEq}) we see that it will go to zero for $\beta=0$. However, our proposed extension given in Eq. (\ref{eq:shinyCrewtherEq}) goes to zero only under both $\beta=0$ and $\alpha\gamma_\alpha =0$. We can attempt to identify which form of the Crewther relation is most accurate for gauge-parameter dependent schemes by evaluating the conformal symmetry breaking term at the roots of the $\beta$-function alone or at fixed points of both the $\beta$-function and $\alpha\gamma_\alpha$. This can be done by evaluating the product of the Adler D function and the Bjorken sum rule at the fixed point. The product is truncated to the same order as the original series. However, due to issues of truncation when evaluated at the fixed point we will not find that $\Deltacsb=0$ exactly; it will only be accurate to the current order in truncation. Therefore in order to get a better idea of the behaviour of this quantity we will consider it at different loop orders to get an idea of its convergence.   

\begin{table}
  \begin{center} 
 \begin{tabular}{ |c||c|c||c|c|} 
 \hline L & $a_{\infty}$&$\alpha_{\infty}$ &$\mathcal{O}(a^3)$&$\mathcal{O}(a^4)$\\  
 \hline 
  2&$0.0033112583$ & $0.0000000000$ &$2.9999991596$ & $3.0000039877$ \\ \hline 
  &$9.1803474173$ & $2.4636080795$ &$1271156.8083213258$ & $17202735.3015072510$ \\ \hline 
  &$0.0032001941$ & $-3.0301823312$ &$2.9999982468$ & $3.0000012469$ \\ \hline 
\hline  
  3&$0.0031177883$ & $0.0000000000$ &$2.9999963264$ & $3.0000001212$ \\ \hline 
  &$0.1279084604$ & $1.9051106246$ &$6.2952539870$ & $10.1893903424$ \\ \hline 
  &$0.0031380724$ & $-3.0274210489$ &$2.9999973439$ & $3.0000001217$ \\ \hline 
 \hline 
  4&$0.0031213518$ & $0.0000000000$ &$2.9999963720$ & $3.0000001843$ \\ \hline 
  &$0.1902883419$ & $0.0000000000$ &$13.5399867931$ & $66.1969134786$ \\ \hline 
  &$0.1162651496$ & $0.5286066929$ &$5.3930704057$ & $11.8942763573$ \\ \hline 
  &$0.0031430130$ & $-3.0273541344$ &$2.9999974127$ & $3.0000002080$ \\ \hline 
 \hline  
  5&$0.0031220809$ & $0.0000000000$ & $2.9999963814$ & $3.0000001972$ \\ \hline 
  &$0.0577103776$ & $0.0000000000$ & $3.2818695828$ & $3.7273436677$ \\ \hline 
  &$0.0031434144$ & $-3.0273765993$ & $2.9999974183$ & $3.0000002151$ \\ \hline 
  &$0.0502252330$ & $-3.8653031470$ & $3.1912609578$ & $3.2787374506$ \\ \hline 
\end{tabular}\label{tab:mMOMCrewtherFP} 
 \end{center}
 \caption{Crewther product evaluated at the fixed points of \cite{13} in $\mMOM$ at different fixed point loop orders $L$. Notation used is from \cite{11}.} \label{tab:CrewtherFPs}
\end{table}

To begin with we will consider Table \ref{tab:CrewtherFPs} which provides the values of the product of the Adler D function and Bjorken sum rule when evaluated at fixed points of the two coupling system which are presented in \cite{13}. The first column provides the loop order the $\beta$-function and $\gamma_\alpha$ are considered to when finding the fixed points, and we define the $\mathcal{O}(a^n)$ columns as the product of the Adler D function and Bjorken sum rule truncated to order $n$ and evaluated at the fixed point. 

To ensure the perturbative nature of the fixed points we have considered them at the top end of the conformal window with sixteen quark flavours, which provides the smallest critical coupling and therefore the most valid perturbative expansion at these points. For the moment we will limit our attention to the Banks-Zaks fixed point \cite{17,18}, which is the closest fixed point to the origin in the Landau gauge found at $a\sim 0.003$, as well as its twin which is infra-red stable and has approximately the same coupling constant but with gauge parameter $\alpha\sim-3$ \cite{19}, as these points should provide the smallest truncation error. We note that if the difference of the product from $d_R=3$ is a truncation error, we only expect improved convergence when both the loop order the fixed point is calculated to and the loop order of $\Deltacsb$ is increased. This is reflected in the table where each value in the $\mathcal{O}(a^3)$ column provides roughly the same accuracy to 3 of $\mathcal{O}(10^{-6})$. Whereas with the exception of the two-loop fixed point in the $\mathcal{O}(a^4)$ column the accuracy is in general $\mathcal{O}(10^{-7})$. Note, as the loop-order in $\Deltacsb$ cannot be increased with the fixed point loop order above this, we do not expect increased accuracy beyond this as we increase the fixed point loop order. This consistency is suggestive of the correct truncation error and therefore fixed points of the two coupling theory provide the roots of $\Deltacsb$ to the order in truncation. This does not mean fixed points of the $\beta$-function alone will not provide similar accuracy and so we will provide comparison briefly.

Before this we should mention the other fixed points further from the Gaussian fixed point. Particularly, we mention that as fixed point loop order is increased, the fixed points move towards the origin and thus the value of $\Deltacsb$ there decreases towards the expected result. However, note that in each case, except at the two loop fixed point, $\mathcal{O}(a^3)$ is smaller than $\mathcal{O}(a^4)$. This is suggestive of truncation error outside of the region of perturbative reliability and therefore these fixed points have only been included in the table for completeness.

\begin{table}
\begin{center} 
 \begin{tabular}{ |c||c||c|c|} 
 \hline 
 L & $a_1$&$\mathcal{O}(a^3)$&$\mathcal{O}(a^4)$\\ \hline 
 2&  $0.0039840637$ & $3.0000169021$ & $3.0000244250$ \\ \hline 
 3&  $0.0037731278$ & $3.0000104128$ & $3.0000164646$ \\ \hline 
 4&  $0.0037925523$ & $3.0000109619$ & $3.0000171393$ \\ \hline 
 5&  $0.0037946540$ & $3.0000110219$ & $3.0000172130$ \\ \hline 
\end{tabular}\label{tab:mMOMCrewtherBFP} 
 \end{center}
  \caption{Crewther product evaluated at the zeros of the $\beta$-function at different loop orders $L$ in the $\mMOM$ scheme with $\alpha=1$ such that $\beta^{\mMOMs}(a_1,1)=0$.} \label{tab:al1CrewtherFPs}
\end{table}

Table \ref{tab:al1CrewtherFPs} provides the values of the roots of the $\beta$-function to different loop orders evaluated in the Feynman gauge $\alpha=1$. In doing this we have picked out the point closest to the origin such that the $\beta$-function disappears, this has roughly the same coupling as the Banks-Zaks fixed point one and so provides the best comparison. By contrast with Table \ref{tab:CrewtherFPs} the values here do not suggest improved convergence between either the $\mathcal{O}(a^3)$ and $\mathcal{O}(a^4)$, nor between the different fixed point loop orders. They remain at a stable $\mathcal{O}(10^{-5})$ from 3. It appears therefore that the fixed point of the two coupling system provides better convergence towards zero in the conformal symmetry breaking term than the fixed points of the single coupling system. 

\begin{figure}
\centering
 \begin{subfigure}{0.7\textwidth}
    \includegraphics[width=\textwidth]{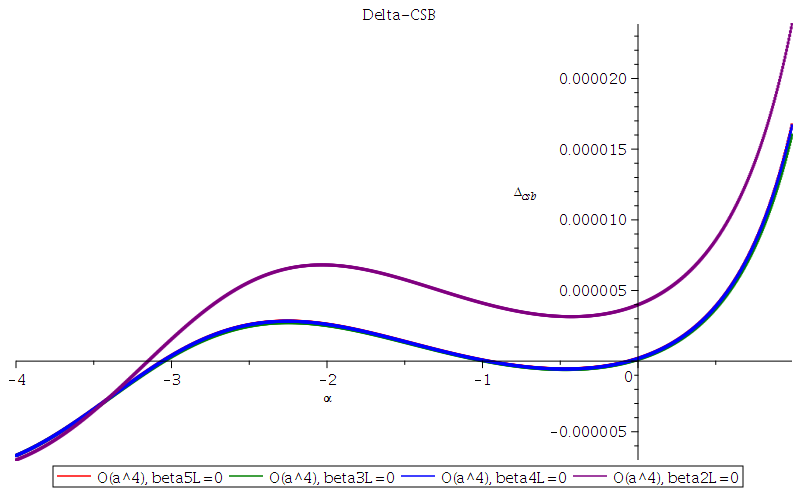}
    \caption{Full Range.}
    \label{fig:first}
\end{subfigure} 
\begin{subfigure}{0.3\textwidth}
    \includegraphics[width=\textwidth]{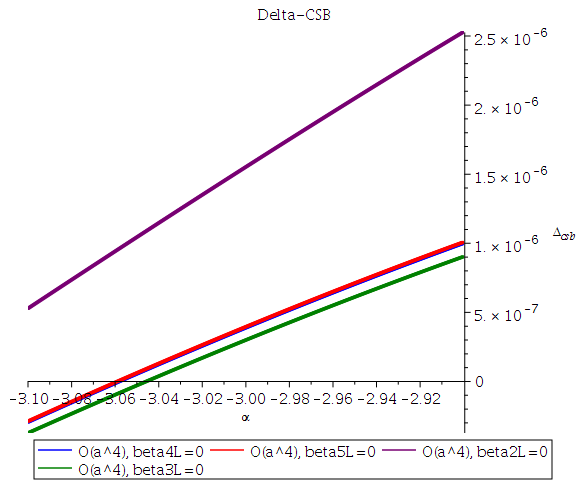}
    \caption{Near $\alpha=-3$.}
    \label{fig:second}
\end{subfigure}
\qquad
\begin{subfigure}{0.3\textwidth}
    \includegraphics[width=\textwidth]{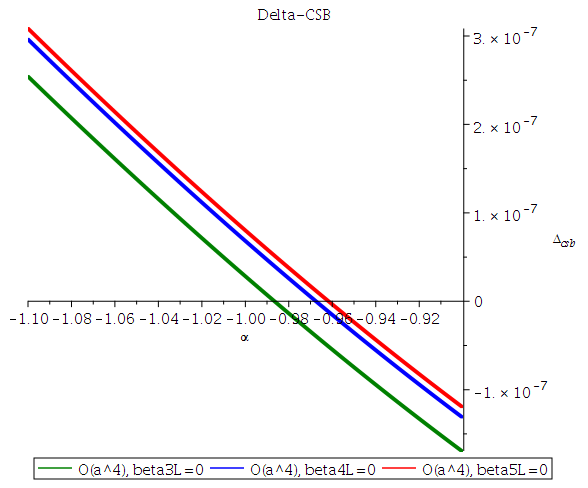}
    \caption{Near $\alpha=-1$.}
    \label{fig:third}
\end{subfigure}
        
\caption{Plots of $\Deltacsb$ calculated to order $\mathcal{O}(a^4)$ in the $\mMOM$ scheme for different $\alpha$ values with $a$ selected as the minimum real, positive value of the coupling constant such that the $\beta_{\mMOM}$ is zero for different loop orders. Note in the first graph the 3L, 4L and 5L graphs are virtually indistinguishable. }
\label{fig:1}
\end{figure}

As a final check on this assumption we have plotted $\Delta_{csb}$ calculated at $O(a^4)$ in Figure 1, at the fixed point closest to the origin; the $x$-axis is $\alpha$. Each line represents the $\beta$-function taken to a different loop order.
 The most obvious feature of the graph is the clear difference between the lines of the two loop fixed point and those of the higher loop orders. Focusing on the higher orders we see a clear cubic structure with the roots at $\alpha\sim0,~-1$ and $~-3$ which are the three gauges of interest highlighted in \cite{9,10}. Figure \ref{fig:second} shows the convergence around $\alpha=-3$, we see it is not absolutely at the anti-Yennie gauge but rather in the vicinity of the Banks-Zaks twin point which was identified in \cite{19} and investigated further in \cite{13}. The Banks-Zaks is the $\alpha=0$ root and $\alpha\sim -1$ appears to be the approximate zero of the $K_\alpha$ series we found in Section \ref{sec:CrewmMOM}.

The $\alpha=-3$ value identified in \cite{9,10} can thus be understood as the point near the Banks-Zaks twin; in fact to leading order in the gauge parameter 
\begin{equation}
\gamma_1(-3) ~=~ \Bigg[-~ \frac{1}{2} \alpha C_A ~+~ \frac{13}{6} C_A ~-~
\frac{4}{3} \Nf T_F\Bigg] \Bigg|_{\alpha=-3}= \beta_0,
\end{equation}
where $\gamma_\alpha(a,\alpha) = \gamma_1(\alpha) a+\mathcal{O}(a^2)$ and $\beta(a,\alpha)=-\beta_0 a^2 +\mathcal{O}(a^3)$. The Banks-Zaks twin infra-red stable fixed point at $\alpha\approx -3$ appears to be the point where the anomalous dimension of the gauge parameter matches that of the coupling constant as we demonstrate in \cite{11}. We can therefore write Crewther's relation in the mMOM scheme to $\mathcal{O}(a^3)$ as
\begin{eqnarray}
\Deltacsb(a,-3)=-K_a^{(0)}\beta_0 a^2 -[K_a^{(0)}\beta_1(-3)+(K_a^{(1)}-3K_\alpha^{(2)}(-3))\beta_0]a^3
\end{eqnarray}
where $K_a^{(i)}$ is the $\mathcal{O}(a^i)$ coefficient of $K_a$ and $K_{\alpha}^{(j)}$ is the $\mathcal{O}(a^j)$ coefficient of $K_\alpha$. Relabelling the leading order $K_a$ term
\begin{eqnarray}
K_a^{(1)}+3K_\alpha^{(2)}(-3)\to K_a^{(1)},
\end{eqnarray}
we see the Crewther product obeys the original relation to $\mathcal{O}(a^3)$ in mMOM. In fact since $\gamma_1(\alpha)$ and $\beta_0$ are scheme independent in the linear covariant gauge this value will be found for all gauge-parameter dependent schemes with this gauge fixing, although analogous values for the gauge parameter have been found for the Curci-Ferrari and Maximal Abelian gauges using equivalent formal relations \cite{11}.

\section{Outlook}\label{sec:outlook}

In this report we have focused on the mMOM scheme as an exemplar for analysing Crewther's relation in gauge-parameter dependent schemes, as the renormalization group functions of this scheme are known to the five-loop level, the key points of the analysis are applicable to other schemes and in other gauge fixing terms provided their $\beta$-function can be related to the $\MSbar$ one by Eq. (\ref{eq:betachange}), as is discussed in more detail in \cite{11}. This highlights the methodology in which we undertook this work: due to truncation each scheme we consider provides an incomplete viewpoint of the underlying structure of the theory, so when we focus too much on a singular scheme we may assume the properties of this particular viewpoint may apply to the theory. In considering properties in generality or else in a variety of schemes we unearth a more complete picture of the theory that may be obfuscated when taking any single finite order calculation.

To summarise, the Crewther relation when considered in a gauge-parameter dependent scheme should account for the running of the gauge parameter and the equation is modified accordingly into the form of Eq. (\ref{eq:shinyCrewtherEq}). We speculate that a natural extension to this relation for systems of $n$ dynamical variables $g_i$ where $i=1,...,n,$ would be:
\begin{eqnarray}
\Deltacsb^s(g^s_i)=\sum_i K_{g_i}^s(g_i^s)\Big(\frac{dg_i^s}{dl}\Big).
\end{eqnarray}
This equation is indicative of the renormalization group equation and so it is worth considering the equation
\begin{eqnarray}
\Deltacsb^s(g_i^s)=\frac{d}{dl}\kappa^s(g_i^s)=\Big(\sum_j \partial_{g_j}^s \kappa(g_i^s) \Big) \Big(\frac{dg_j^s}{dl}\Big)\label{eq:kappacsb}
\end{eqnarray}
where $\partial_{g_j}^s=\frac{\partial }{\partial g_i^s}$. For our two-coupling theory this reduces to
\begin{eqnarray}
\Deltacsb^s(a_s,\alpha_s)=\Big(\partial^s_{a} \kappa^s(a_s,\alpha_s) \Big)\beta^s(a_s,\alpha_s)+\Big(\partial_{\alpha}^s \kappa^s(a_s,\alpha_s) \Big)\alpha_s\gamma^s_\alpha(a_s,\alpha_s).
\end{eqnarray}
Investigating this for the Crewther relation in our two-coupling theory we find that provided the $\kappa$ series in $\MSbar$ is gauge-parameter independent, as one would expect for a perturbative series in this scheme,  then the above relation reduced trivially to Eq. (\ref{eq:CrewtherEq}) with 
\begin{eqnarray}
K^{\MSbars}_a(a_{\MSbars})=\partial_a^{\MSbars}\kappa^{\MSbars}(a_{\MSbars}).
\end{eqnarray}
Integrating this equation with respect to the coupling constant can then be used to define $\kappa$ in the $\MSbar$ scheme. If  Eq. (\ref{eq:kappacsb}) holds then $\kappa$ is a scheme independent quantity, therefore under a scheme transformation $\partial_a^{\MSbars}\kappa^{\MSbars}$ will transform in the same way as was found for $K_a^{\MSbars}$ and so if the relation holds in $\MSbar$ it should hold in all other schemes. This can be shown by directly applying a scheme transformation to the above equations to ensure consistency. The ambiguity laid out in Eq. (\ref{eq:CrewtherAmbiguity}) could then be understood as resulting from shifting $\kappa$ by a conformally invariant quantity.

Beyond this, our analysis of the Crewther relation is indicative of the requirement for the modification of a wider treatment of running of gauge-parameter dependent schemes. For example, in any instance which codifies the running of theory, or else provides a decomposition of a measurable, in terms of the $\beta$-function will likely need to be extended for gauge-parameter dependent schemes to include the running of the gauge parameter, e.g. \cite{5,20, 21}.

\acknowledgments This work was carried out with the support of an EPSRC Studentship EP/R513271/1 
(RHM) and the STFC Consolidated Grant ST/T000988/1 (JAG). For the purpose of open access, the authors have applied a Creative 
Commons Attribution (CC-BY) licence to any Author Accepted Manuscript version arising.

\end{document}